\documentclass[sigconf, 10pt]{acmart}
\usepackage{subcaption}
\usepackage{enumitem}
\usepackage{xcolor}
\usepackage{tcolorbox}
\usepackage{xurl}
\usepackage{multirow}
\usepackage{colortbl}
\usepackage{tikz}

\usepackage{comment}
\renewcommand\footnotetextcopyrightpermission[1]{} % removes footnote with conference info
\setcopyright{none}
\acmYear{2025}\copyrightyear{2025}
\setcopyright{rightsretained}
\acmConference[PRIMES '25]{ACM IMC Workshop of Policy-Relevant Internet Measurements and Experimentation (PRIMES)}{October, 2025}{Madison, Wisconsin}
\acmBooktitle{ACM IMC Workshop of Policy-Relevant Internet Measurements and Experimentation (PRIMES), October, 2025, Madison, Wisconsin}

\begin{document}

\title{Expanding Access, Exposing Risk:\\ A Short Study of Exposed Starlink Hosts}


\author{Omar Elamri*}
\thanks{*Equal contribution}
\affiliation{%
  \institution{UCLA}
  \city{Los Angeles}
  \country{USA}
}
\email{omar@cs.ucla.edu}

\author{Isaac-Neil Zanoria*}
% \authornote{Equal contribution}
\affiliation{%
  \institution{UCLA}
  \city{Los Angeles}
  \country{USA}
}
\email{zanoria@ucla.edu}

\author{Jacob Zhi*}
% \authornote{Equal contribution}
\affiliation{%
  \institution{UCLA}
  \city{Los Angeles}
  \country{USA}
}
\email{zhi@cs.ucla.edu}

\author{Ben Du}
\affiliation{%
  \institution{UCLA}
  \city{Los Angeles}
  \country{USA}
}
\email{c4du@ucla.edu}

\author{Liz Izhikevich}
\affiliation{%
  \institution{UCLA}
  \city{Los Angeles}
  \country{USA}
}
\email{lizhikev@ucla.edu}


% \begin{abstract}
%  % \input{abstract}
% \end{abstract}

%    <concept>
%        <concept_id>10003033.10003083.10003014</concept_id>
%        <concept_desc>Networks~Network security</concept_desc>
%        <concept_significance>500</concept_significance>
%        </concept>
%    <concept>
%        <concept_id>10003033.10003079.10011704</concept_id>
%        <concept_desc>Networks~Network measurement</concept_desc>
%        <concept_significance>500</concept_significance>
%        </concept>
%  </ccs2012>

% \ccsdesc[500]{Networks~Network security}
% \ccsdesc[500]{Networks~Network measurement}

%
% Keywords. The author(s) should pick words that accurately describe
% the work being presented. Separate the keywords with commas.
% \keywords{}

\maketitle

\section{Introduction}

Low Earth Orbit (LEO) satellite networks are reshaping global Internet access by connecting regions that have long lacked reliable terrestrial infrastructure. Among them, Starlink has seen explosive growth—reporting over 7 million users in 2025 \cite{starlink2025}, up from just 2 million in 2023 \cite{starlink2023}—primarily targeting rural and underserved areas. This rapid expansion brings millions of new hosts online, many of which may differ in security posture from those behind traditional broadband networks.
\looseness=-1

In this paper, we present a measurement-driven analysis of the security characteristics of Starlink-connected hosts and uncover several concerning trends. We find that
Starlink hosts are more likely to run outdated or vulnerable operating systems and network protocols than non-Starlink hosts.
These disparities are not uniform globally—regions like Latin America, Southeast Asia, and Eastern Europe show disproportionately higher risk.

Our findings raise important questions for the Internet measurement and policy communities: How can we reduce the security risks associated with newly connected populations? What role should ISPs, manufacturers, and governments play in promoting safer defaults and practices? We hope this work can serve as a basis for discussion on how to design technical and policy interventions that ensure connectivity growth also strengthens the global Internet’s resilience.

\section{Dataset \& Methodology}

We collect IPv4 and IPv6 services exposed behind Censys~\cite{censys} on April 7th, 2025. These scans provide general information about responsive hosts (such as their IP address, autonomous system, and geolocation) as well as the OS and service-related information obtained through protocol banners. 

%\textbf{Identifying starlink and non-starlink hosts.} 

We identify services hosted within Starlink's autonomous system as Starlink hosts, and all others as non-Starlink hosts. This yields approximately 33,000 Starlink hosts. To ensure a balanced comparison, we reduce our original set of ~230 million non-Starlink hosts by randomly sampling up to 10K hosts per country, resulting in a dataset of 1.78 million non-Starlink hosts. This slightly exceeds the maximum number of Starlink hosts observed in a single country.

%\textbf{Identifying vulnerabilities associated with Starlink hosts.} 

We obtain information about recent CVE's--those between 2020 and 2025--through the NIST's National Vulnerability Databse program.
The majority of the CVE descriptions in this database include a range of affected operating system and software versions.
We join this dataset with both Censys datasets on a description of the host system using a wildcard match on its Common Platform Enumeration (CPE) string to obtain a join table of CVE-host pairs.

\section{Results}

\begin{figure}
    \centering
    \includegraphics[width=0.9\linewidth]{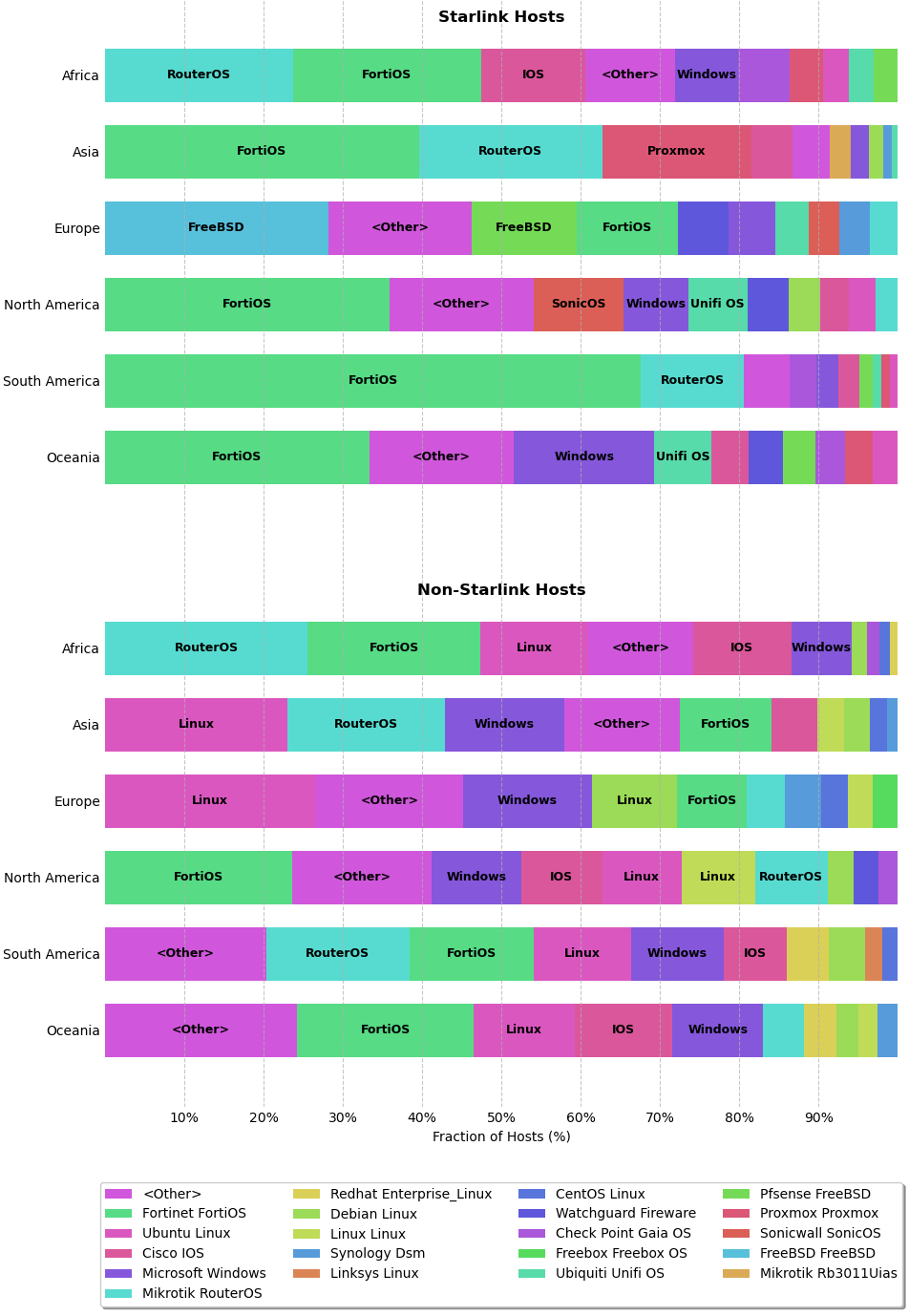}
    \caption{Starlink hosts observed in the Censys dataset are primarily routers compared to non-Starlink hosts.}
    \label{fig:osdistribution}
\end{figure}

A significant fraction of exposed Starlink hosts are routers, with the type and frequency of their operating systems varying significantly across continents.
%Figure~\ref{fig:osdistribution} shows that while a significant fraction of operating systems seen on Starlink are router operating systems, the type and frequency of these operating systems shift dramatically between continents. 
South America sees 70\% of exposed hosts running Fortinet FortiOS, while Europe sees less than 20\% of the same operating system.
Figure~\ref{fig:osdistribution} shows that these geographic variations are especially apparent when compared to the (relatively even) distribution of non-Starlink hosts. For example, FortiOS takes up 15.72\% market share in South America for non-Starlink hosts.

\begin{figure}
    \centering
    \includegraphics[width=0.9\linewidth]{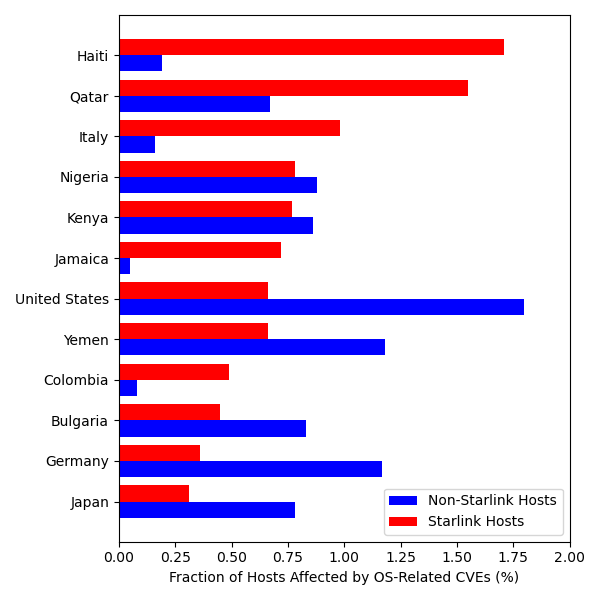}
    \caption{12 most common countries with Starlink-connected hosts affected by an operating system CVE versus non-Starlink hosts. Data is normalized by the total number of hosts.}
    \label{fig:os-cve-disparity}
\end{figure}

Starlink hosts show significantly higher rates of operating system vulnerabilities compared to non-Starlink hosts, with notable geographic disparities. As shown in Figure~\ref{fig:os-cve-disparity}, countries with the highest OS CVE rates for Starlink hosts do not necessarily align with those for non-Starlink hosts. Notably, Haiti experiences a nine-fold increase in per-capita OS vulnerabilities on Starlink-connected hosts, while Colombia sees a six-fold increase.

Starlink hosts are more likely to run outdated and insecure network protocols than non-Starlink hosts, with this pattern varying significantly across countries. To quantify this difference, we compute a Starlink insecurity ratio for each country—defined as the fraction of Starlink hosts using outdated protocols (including old TLS versions, weak SSH cipher suites, and legacy SMBv1) divided by the corresponding fraction among all hosts.

\begin{figure}
    \centering
    \includegraphics[width=0.9\linewidth]{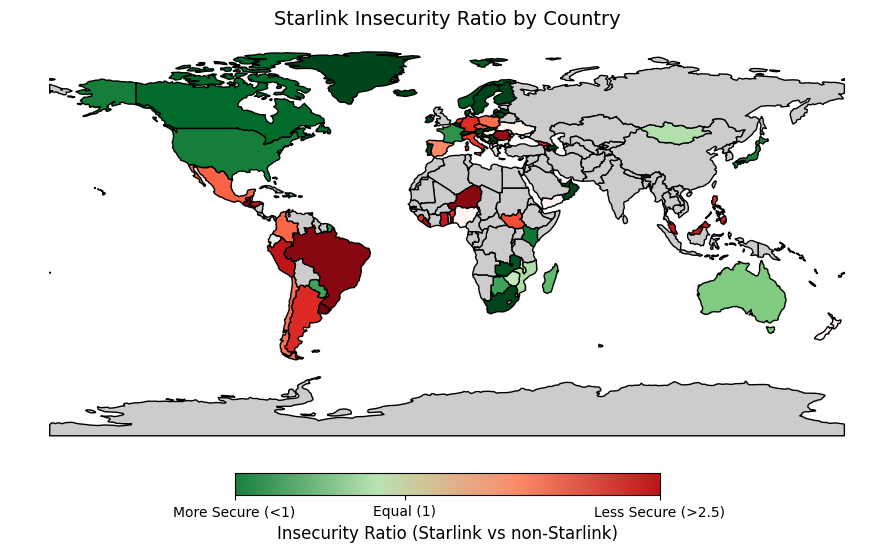}
    \caption{Starlink hosts in South America, west Africa, and eastern Europe contribute to disportionately to overall hosts that runs outdated network protocols.}
    \label{fig:prot-distr}
\end{figure}

Starlink overrepresents insecure protocol usage in regions such as Central and South America, Eastern Europe, and Southeast Asia.
Figure~\ref{fig:prot-distr} shows the global distribution of this ratio. Countries shaded green have a ratio below 1, indicating Starlink hosts are comparatively more secure, while those in red have a ratio above 1, reflecting higher insecurity among Starlink hosts.

\section{Discussion}

Our sampled results point to a critical and timely policy challenge: how to ensure that the rapid expansion of connectivity—particularly through satellite ISPs like Starlink—does not come at the expense of global Internet security. 
The disparities we observe in OS types, vulnerability rates, and outdated protocol use across regions suggest that the attack surface is not only growing, but also unevenly distributed. 
This asymmetry may correlate with regional differences in technical capacity, regulation, and user awareness—factors that policy and community efforts could potentially address.

As outdated or vulnerable devices become easy targets for botnets and exploitation, the Internet community could coordinate efforts to promote security awareness and stronger defaults in newly connected regions to ensure that expanded access does not come at the cost of resilience.

%
% The acknowledgments section is defined using the "acks" environment
% (and NOT an unnumbered section). This ensures the proper
% identification of the section in the article metadata, and the
% consistent spelling of the heading.
% \begin{acks}

% \end{acks}

\bibliographystyle{ACM-Reference-Format}
%\balance
\bibliography{references}

\appendix

\end{document}